\documentclass[fleqn,usenatbib]{mnras}

\usepackage{newtxtext,newtxmath}

\usepackage[T1]{fontenc}
\usepackage{ae,aecompl}
\usepackage{pdflscape}
\usepackage{array}

\usepackage{graphicx}	
\usepackage{amsmath}	
\usepackage{amssymb}	
\usepackage[font=small,labelfont=bf]{caption} 

\title[Small scale star formation as revealed by VVVX galactic cluster candidates]{Small scale star formation as revealed by VVVX galactic cluster candidates.}
\author[J. Borissova et al.]{
J. Borissova,$^{1,2}$\thanks{Based on observations gathered with VIRCAM at the ESO VISTA telescope, as part of observing programs 198.B-2004.}
R. Kurtev,$^{1,2}$
N. Amarinho,$^{3}$
J. Alonso-Garc\'ia,$^{4,2}$
S. Ram\'irez Alegr\'ia,$^{4},$
\newauthor
S. Bernal,$^{1,2}$
N. Medina,$^{1,2}$
A.-N. Chen\'e,$^{5}$ 
V.~D. Ivanov,$^{6}$
P.W. Lucas,$^{7}$
and
D. Minniti$^{8}$ 
\\
$^{1}$Instituto de F\'isica y Astronom\'ia, Universidad de Valpara\'iso, Av. Gran Breta\~na 1111, Playa Ancha, Casilla 5030, Chile.\\
$^{2}$Millennium Institute of Astrophysics (MAS), Santiago, Chile.\\
$^{3}$Laborat\'orio Nacional de Astrof\'isica (LNA) - Brazil \\
$^{4}$Centro de Astronom\'{i}a (CITEVA), Universidad de Antofagasta, Avenida Angamos 601, Antofagasta, Chile \\
$^{5}$NOIRLab international Gemini Observatory, Northern Operations Center, 670 A'ohoku Place, Hilo, HI 96720, USA\\
$^{6}$European Southern Observatory, Karl Schwarzschildstr. 2, D-85748 Garching bei M\"unchen, Germany\\
$^{7}$Centre for Astrophysics, University of Hertfordshire, College Lane, Hatffeld, AL10 9AB, UK.\\
$^{8}$Departamento de F\'isica, Facultad de Ciencias Exactas, Universidad And\'es Bello, Av. Fernandez Concha 700, Las Condes,
Santiago, Chile.\\}
\date{Accepted XXX. Received YYY; in original form ZZZ}

\pubyear{2020}

\begin{document}
\label{firstpage}
\pagerange{\pageref{firstpage}--\pageref{lastpage}}
\maketitle

\begin{abstract}
We report a search and analysis of obscured cluster candidates in the ``VISTA Variables in the Via Lactea eXtended (VVVX)'' ESO Public Survey area encompassing the region between $229^\circ.4 < l < 295^\circ.2 $ and $-4^\circ.3 < b < 4^\circ.4$ of the southern Galactic disk. We discover and propose 88 new clusters. We improve the completeness of the embedded cluster population in this region, adding small size (linear diameters of 0.2-1.4 pc) and relatively far objects (heliocentric distance between 2 and 4 kpc) to existing catalogues. Nine candidates are proposed to be older open cluster candidates. Three of them (VVVX\,CL\,204, \,207, \,208) have sufficient numbers of well-resolved stellar members to allow us to determine some basic cluster parameters. We confirm their nature as older, low-mass open clusters. Photometric analysis of 15 known clusters shows that they have ages above 20\,Myr, and masses below 2000\,M$_{\odot}$: in general, their proper motions follow the motion of the disk. We outline some groups of clusters, most probably formed within the same dust complex. Broadly, our candidates follow the network of filamentary structure in the remaining dust. Thus, in this part of the southern disk of the Galaxy, we have found recent star formation, producing small size and young clusters, in addition to the well known, massive young clusters, including NGC\,3603, Westerlund\,2 and the Carina Nebula Complex.
\end{abstract}

\begin{keywords}
Galaxy: open clusters and associations -- Galaxy: disk --Infrared: stars
\end{keywords}



\section{Introduction}

The investigation of open star clusters has been very useful for the exploration of the disk of our own Milky Way.
They are excellent laboratories not only for stellar evolution theory, but also to test the star formation history and the structure of the Galactic disk (e.g. \citealt{Moi10}). 

This is an active area of research where good progress is being made. In the optical regime, the Gaia mission has enabled discovery of hundreds of cluster candidates, reviewed known clusters and distinguished asterisms among objects previously reported as stellar clusters (e.g. \citealt{cantat-gaudin18,cantat-gaudin19} and \citealt{cantat-gaudin20};  \citealt{2019A&A...628A..45G}). The near-infrared data allow access to more embedded and younger objects, completing the view obtained through optical data. The early discoveries done with the 2\,MASS survey (e.g. \citealt{bica03,dutra03,froebrich07}) have been recently complemented with VVV survey data (e.g. \citealt{Bor11,Bor14,Bor18,Sol14,Bar15,Iva17}; and references therein) through a variety of techniques. The future searches with Gaia DR3 and the LSST are also expected to yield large numbers of open clusters  \cite[e.g. The LSST Science Book,][]{science_book}.

The characterization of these open cluster candidates is often tricky, as they are located at low Galactic latitudes where not only the stellar density is high, but also the interstellar extinction and reddening are problematic. 
It is not surprising that different approaches have been taken to measure the physical parameters for these clusters, often using vastly different datasets. 
Fortunately, the large datasets now available allow us to shift from the analysis of single objects to the study of large samples like the ones produced now by Gaia and VVV, and also LSST in the future. 
In particular, we can now try to trace the star formation patterns at different scales across the Galactic disk. 
 
The near-IR offers advantages over optical wavelengths for these kinds of studies because it is less sensitive to reddening and extinction. 
Using the near-IR we can see  for example, young, obscured clusters farther away, deeper into the Galactic plane, and can also  measure more accurate distances, that are less affected by reddening uncertainties. 
There is a disadvantage, however: the background contamination in the near-IR is generally much larger than in the optical, because at shorter wavelengths the same extinction helps to reduce the numbers of distant stars that can be detected. 
Thus,  the optical color magnitude diagrams of nearby clusters often appear cleaner than the more distant ones that we are finding in the near-IR. Therefore, both optical and near-IR studies are worthwhile and complementary. 

In this work we take advantage of the near-IR VVVX survey, searching for distant ($1.0<D({\rm kpc})<10.0$) young clusters at very low Galactic latitudes ($-5 ^\circ < b < +5^\circ$)  and use them to trace small scale star formation across their fields. This is the fourth paper in our effort to study the obscured cluster population in the MW \citep{Bor11,Bor14,Bor18}, using the ``VISTA Variables in the V\'{\i}a L\'actea'' (VVV) and the new ``VVV eXtended'' (VVVX) surveys that
cover $\sim$1700\,deg$^2$ of the inner MW bulge and southern disk in $ZYJHK_S$ bands \citep{Min10,Sai12}.
The VVV/VVVX images have a pixel size of $0.34\arcsec$, smaller than 2\,MASS \citep[Two Micron All Sky Survey;][]{Skr06}, GLIMPSE \citep[Galactic Legacy Infrared Mid-Plane Survey Extraordinaire;][]{Ben03} and WISE \citep[Wide-field Infrared Survey Explorer,][]{Wri10} -- especially important for searching of small/compact cluster candidates. This allows us to investigate the formation of such objects and to connect them with the dust distribution, obtained with the mid-IR observations.

\begin{figure}
\begin{center}
\includegraphics[width=\columnwidth]{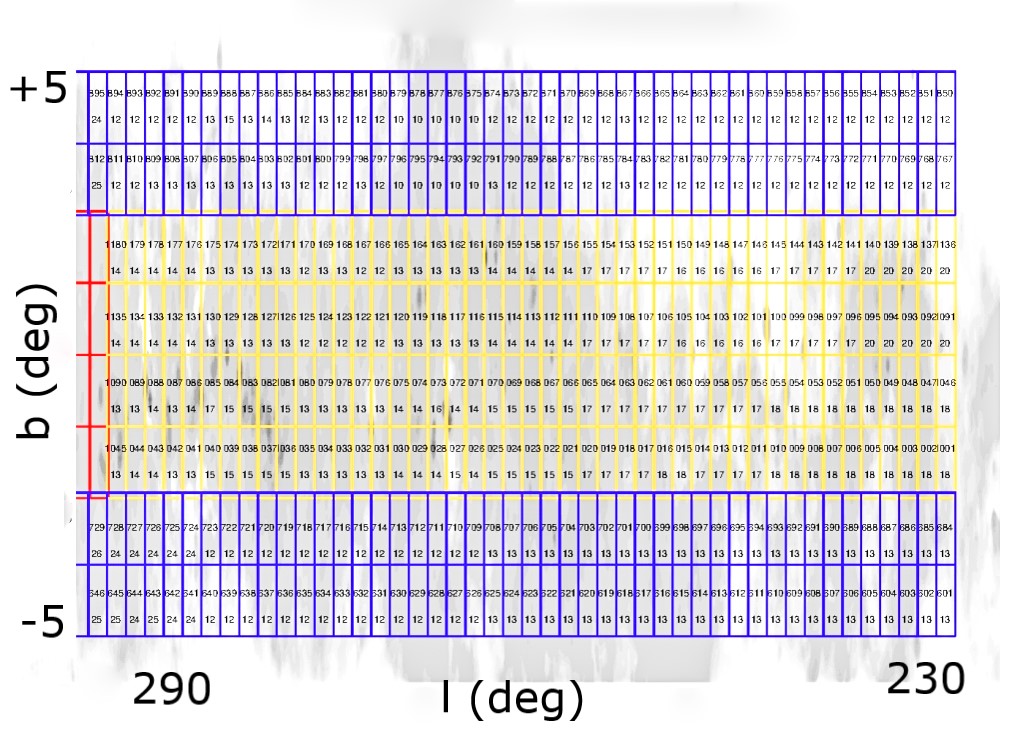}
\end{center}
\caption{The VVVX Survey area, investigated in this work. The blue and yellow squares show the VVVX outer disk fields. The numbers in each tile are its identification number (top) and the number of available $K_S$ epochs, as of May 2, 2019.}
\label{tiles_id}
\end{figure}

Here we expand our search into regions recently observed by the VVVX survey (Fig.\,\ref{tiles_id}): two slices away from the plane in the southern disk ($65\degr \times 4\degr$,
tiles from e601 to e729 and from e767 to e895; lower and upper part, marked in blue) and an extension along the mid-plane ($65\degr \times 4\degr$, tiles from e1001 to e1180; shown in yellow). The analyzed area encompass the region between $229\fdg4 < l < 295\fdg2 $ and $-4\fdg3 < b < 4\fdg4$. A description of the survey data is included in Section~\ref{survey}. Sections~\ref{search} and \ref{validation} detail the candidate detection methods and validation. Section~\ref{known} summarizes our review of known clusters in the search area, reported by \citet{Bica19}.  The spatial distribution of clusters with distance determination, identifying several groups and structures, is discussed in Section \ref{spacial}.

\section{Survey data}\label{survey}

The VVVX survey was launched in 2016 as an extension of the completed VVV survey in order to enhance its legacy value. The VVVX provides a spatial coverage from $l$=230$^\circ$ to $l$=20$^\circ$ (7$^{\rm h}$\,$<$\,$\alpha$\,$<$\,19$^{\rm h}$). The data were obtained with the 4.1-meter ESO VISTA telescope \cite[Visual and Infrared Survey Telescope for Astronomy;][]{Eme06} located at Cerro Paranal, Chile, with the 16-detector VIRCAM \citep[VISTA Infrared CAMera;][]{Dal06}. It has a $\sim$1$\,\times$\,1.5\,deg$^2$ field of view, works in the 0.9-2.5\,$\mu$m wavelength range, and has a pixel scale of 0.34\,arcsec\,px$^{-1}$.

The data are reduced with the VISTA Data Flow System \cite[VDFS;][]{Irw04,Eme04} at the Cambridge Astronomical Survey Unit\footnote{\url{http://casu.ast.cam.ac.uk/}} (CASU). Processed images and photometric catalogs are available from the ESO Science Archive\footnote{\url{http://archive.eso.org/}} and from the VISTA Science Archive\footnote{\url{http://horus.roe.ac.uk/vsa/}} \citep[VSA;][]{Cro12}. 

A single VIRCAM image, called pawprint, contains large gaps; six pawprints taken in a spatial offset pattern must be combined to fill them in, obtaining a contiguous image, called tile. Each point of the sky is imaged at least twice within a tile, except of the outermost edges that are imaged once. The VVVX pawprints and tiles are aligned along $l$ and $b$. The total effective exposure time of the tiles are: 8\,sec in $K_S$ (for a single epoch), 24\,sec in $H$ and 60\,in $J$ band. 

In order to obtain the best results in such a crowded environment, we extract our own point spread function (PSF) photometry on the different regions of interest.  This procedure is described in detail by \cite{Alo18} and \cite{Bor11,Bor14} and enhances the depth of CASU aperture photometry, typically by 1-2 mag, along with the completeness of our data. The magnitudes, calibrated in the VISTA system, were transformed to the 2\,MASS standard system, and the saturated stars (usually with $K_S\leq11.5$ mag, depending on crowding) were replaced with 2\,MASS photometry, following \citet{Alo18}.


\section{Cluster Search and candidate selection} \label{search}

We searched cluster candidates by means of visual inspection, as in our previous works \citep{Bor11,Bor14,Bor18}. 
The observed images were retrieved from the CASU database. Initially the $K_S$ tiles were visually inspected, creating a preliminary list of candidates. The main criterion was the presence of a local overdensity with respect to the surrounding area and (or) visually connected with any surrounding nebulosity. Then, composite $JHK_S$ color images were created. We verified the compact nature of the object and required at least 5-6 stars with similar colors to be clustered at the objects' center. This method was preferred over various automatic algorithms that count stars, in order to include objects that are not fully resolved into members.

We have identified 88 new cluster candidates. They are listed in Table\,\ref{candidates}. The first column gives the identification, followed by the equatorial coordinates of the center determined by eye, eye-ball measured apparent cluster radius in arcmin, the name of the corresponding VVVX tile, distance to the candidate in kpc (if available), {\it Herschel} SPIRE 500\,$\mu$m measured flux in Jy, {\it WISE} W4 22\,$\mu$m measured flux in Jy, dust mass in units of solar masses (both variables are described below in the text) and any nearby associated (within the visual cluster diameter) objects such as IR sources, YSOs, \ion{H}{ii} regions, dense cores, etc. The cluster radii were measured by eye on the VVVX $K_S$ tiles. To do this, the area around cluster candidates is smoothed and the stellar density contours are over-plotted with the lowest contour corresponding to the stellar density of comparison field. 

\begin{figure*}
\begin{center}
\includegraphics[width=14cm]{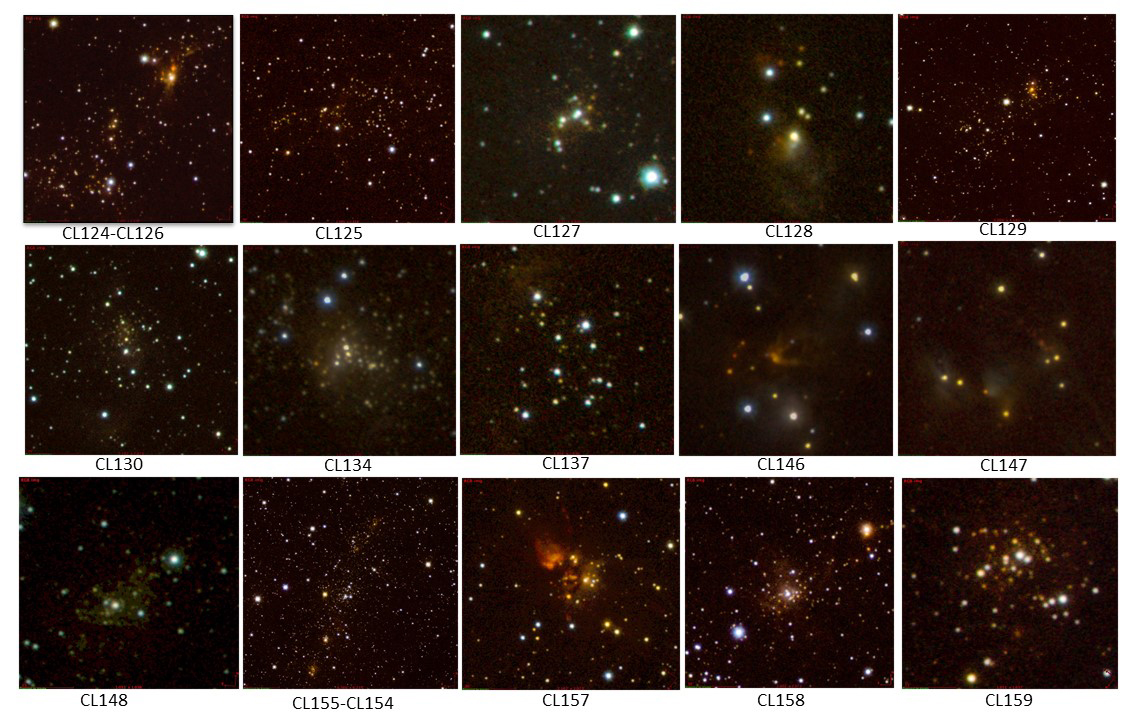}
\includegraphics[width=14cm]{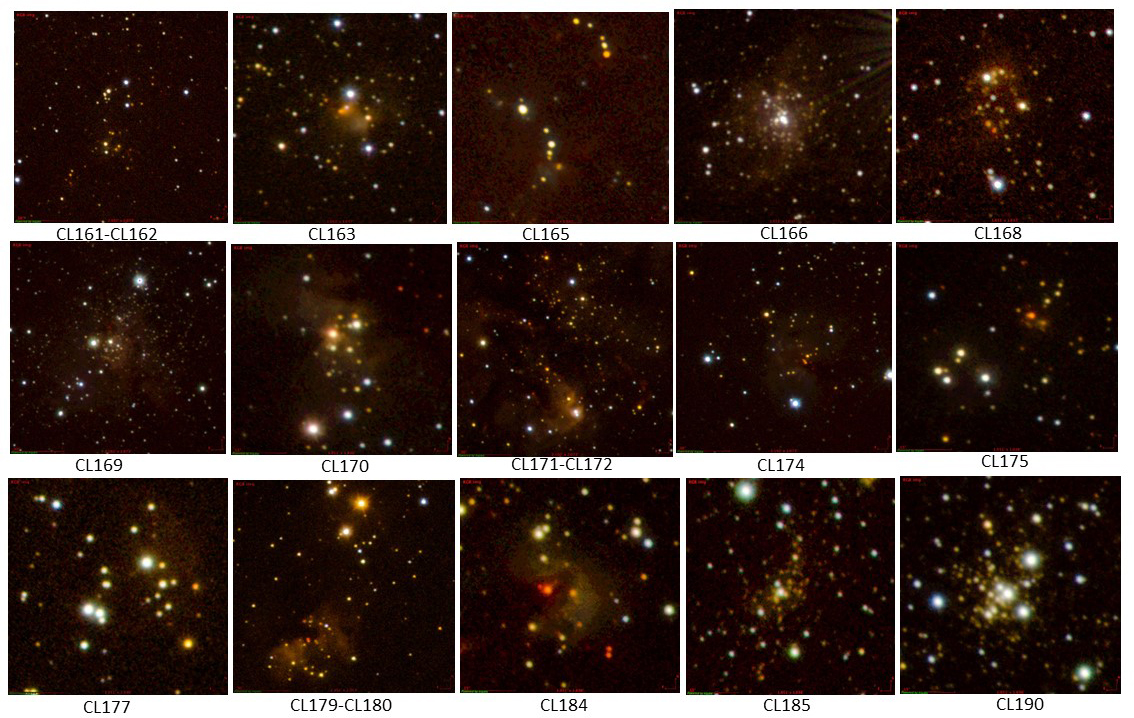}
\end{center}
\caption{VVVX $JHK_S$ color images for a subset of the newly found cluster candidates. The field of view is typically 1.5x1.5 arcmin; North is up, East is to the left.}
\label{rgb}
\end{figure*}

\begin{figure*}
\begin{center}
\includegraphics[width=8cm]{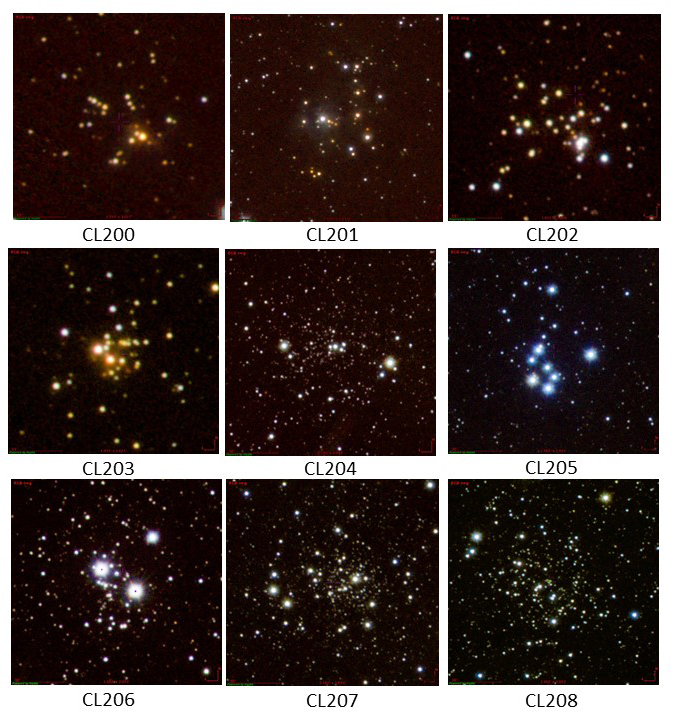}
\end{center}
\caption{VVVX $JHK_S$ color images of newly found cluster candidates. The field of view is typically 2.5x2.5 arcmin; North is up, East is to the left.}
\label{VVVX_oc}
\end{figure*}

Figures\,\ref{rgb} and \ref{VVVX_oc} show the VVVX $JHK_S$ color images of some of them for illustration.

\section{Validation and preliminary classification}\label{validation}

\subsection{Obscured cluster candidates}

According to \citet{Asc18} ``An embedded cluster is one that is still enshrouded in its natal molecular cloud. It is typically not (fully) observable at optical wavelengths due to the heavy
obscuration caused by the dust grains in the cloud, but it can be seen in the near infrared, where young stars emit significantly and the dust is more transparent''. As pointed out in the same review, the upper limit of the age of such objects is 5-10 Myr. 

To investigate the probability of physical connection between our cluster candidates and any remaining dust and gas in their vicinities, we overplot the cluster candidates on far-IR color images. Some good tracers of cold dust are {\it Herschel} PACS \citep[70\,$\mu$m -- blue, 160\,$\mu$m -- red;][]{Pog10} and SPIRE \citep[250\,$\mu$m -- blue; 350\,$\mu$m -- green; and 500\,$\mu$m -- red;][]{Gri10}. These are taken from the  archival HiGal survey \citep{molinari10}. Fifty four (61 \%) of our 88 candidates have Herschel data, the rest are projected outside of the field of view. We performed aperture photometry on the {\it Herschel} SPIRE 500\,$\mu$m normalized image, chosen because of the lower mean temperature of the dust (18.7\,K, \citealt{Zhu14}). The diameter of aperture was determined on the base of flux contours. The typical flux density uncertainty is 15\%. The mass of the dust is calculated following the calibration of \citet[][see Eqn.\,1]{Par12}, assuming that the main contribution of the continuum emission come from the dust. Thirty candidates have SPIRE 500\,$\mu$m fluxes with the 3$\sigma$ values above the local field and are probably still associated with their cold dust.  Most of them coincide with the brightest core of the extended dust emission. They are listed in Table\,\ref{candidates} together with their corresponding dust mass. 
 
We also verified the projection of our candidates on the {\it allWISE} color composite  (W4 -- red; W2 --  green; W1 --  blue; \citealt{Wri10}) image. Table\,\ref{candidates} lists the  W4 (22\,$\mu$m) band fluxes with the 3$\sigma$ values above the local field. The typical flux density uncertainty is around 10\%. Only 20 objects can not be connected with any dust emissions in this band. 

\begin{figure*}
\begin{center}
\includegraphics[width=15cm]{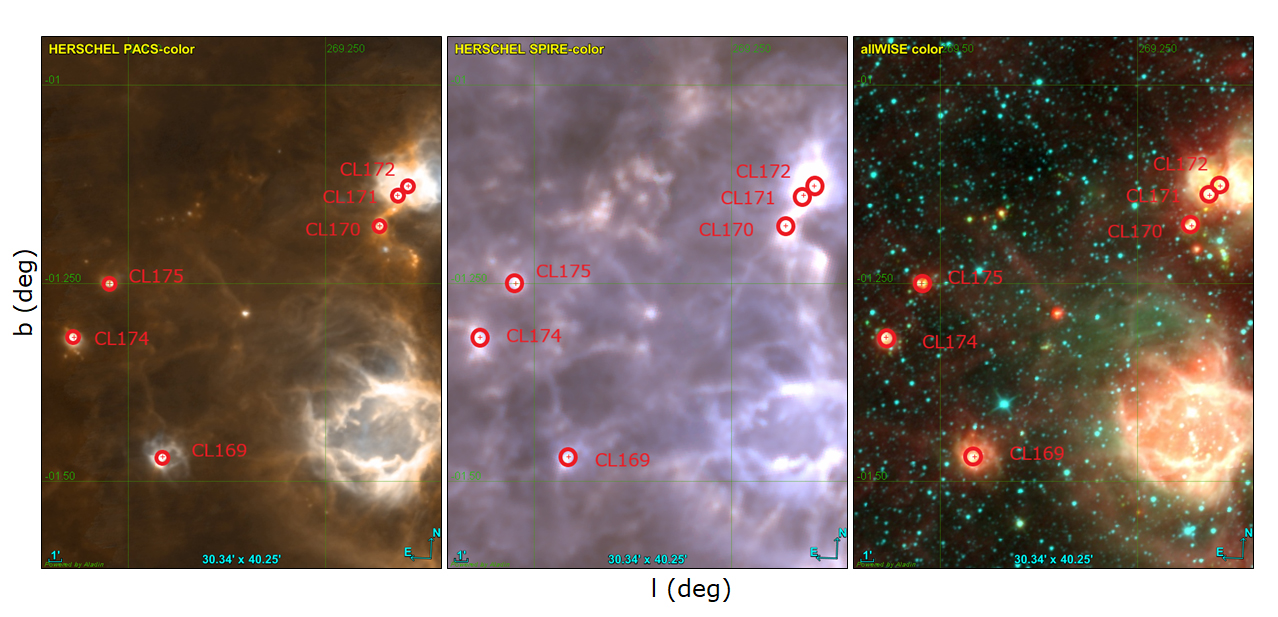}
\end{center}
\caption{{\it Herschel} PACS and SPIRE and {\it allWISE} color compositions with overplotted CL\, 169, 171, 172, 174 and 175. The field of view is 30 x 40 arcmin; the North is up, East is to the left.}
\label{her}
\end{figure*}

Figure\,\ref{her} shows the {\it Herschel} and {\it allWISE}  images with some candidates overplotted for illustration.

Thus, the emission clumps in the dust continuum in {\it Herschel} and {\it allWISE}  suggested that 71 (80 \%) of our objects can be classified as obscured, possibly still embedded in their natal dust.

In a search for indicators of youth, such as young stellar objects (YSOs), masers, IR sources and \ion{H}{II} regions, we cross-identified the objects in our sample with the SIMBAD astronomical database\footnote{\url{http://simbad.u-strasbg.fr/simbad/}} \citep{Wen00}.  We found indicators of youth for 63 of our candidates (see Table\,\ref{candidates} for comments on individual objects). They can be classified as young, with ages of 5-10\,Myr or less.

\begin{figure*}
\begin{center}
\includegraphics[width=16cm]{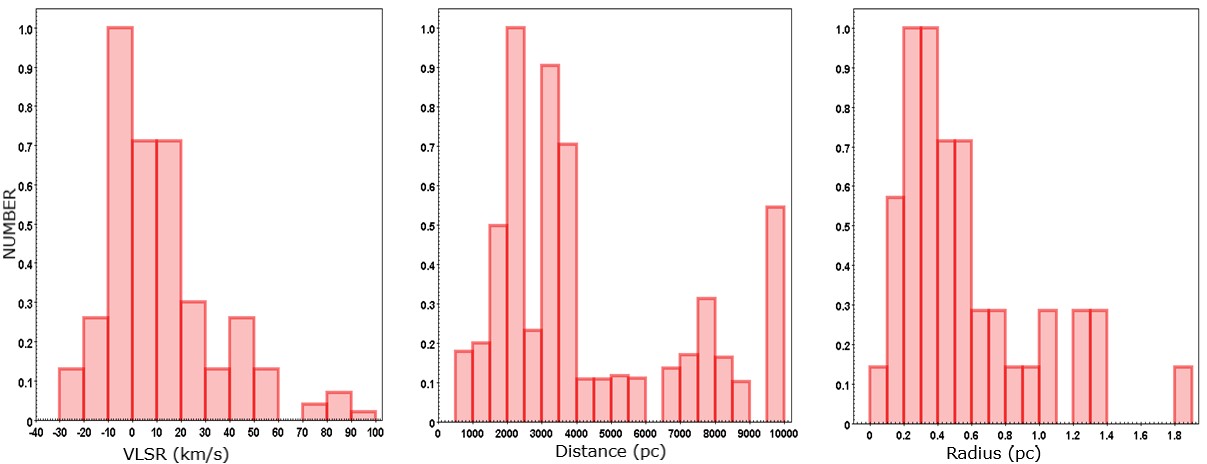}
\end{center}
\caption{Normalized distributions of (left to right): VLSR (km\,s$^{-1}$), kinematic distance (pc) and linear radius (pc) of the cluster candidates.}
\label{dis}
\end{figure*}

Radial velocities in the local system of rest (VLSR) are available in the literature for 42 of our candidates. Using the \citet{Wen18} tool\footnote{\url{https://www.treywenger.com/kd/index.php}}, we calculated their kinematic distances, taking into account the VLSR uncertainties. The tool uses the \citet{Reid14} MW rotation curve with updated solar motion parameters and evaluates the errors with 10\,000 Monte Carlo realizations. It seems optimal for our small size and young clusters. The results are listed in Table\,\ref{candidates} and the normalized histogram distributions of VLSRs, distances and cluster radii are shown in Fig.\,\ref{dis}. Most of the candidates are located at distances 2-4\,kpc from the Sun and have linear diameters of 0.2-1.4\,pc.

\subsection{Open cluster candidates}\label{open_cl}

Nine clusters (listed in Table\,\ref{candidates} from CL\,200 to CL\,208) are not projected close to any nebulosity, but show up as clear overdensities with respect to the surrounding field. They are resolved or semi-resolved. Here, the validation process includes the analysis of the CMDs \citep[see][for more details]{Bor18}. Distances from {\it Gaia} Data Release 2 \citep[{\it Gaia} DR2;][]{gaia-collaboration2018} and \cite{Bailer18} were considered, when available. However, most objects are sparsely populated and only three of them (CL\,204, CL\,207 and CL\,208) have a high enough number of possible member stars for estimation of any fundamental cluster parameters. Deeper observations with higher angular resolution are needed to reveal the nature of the remaining six open cluster candidates.

For the analysis of VVVX CL\,204, CL\,207 and CL\,208, the first step was to construct their CMDs from the PSF photometry within 5$\times$5\,arcmin area around the cluster center.  Then, the \cite {Bonatto19} tools $\it {CLEAN_{phot}}$ and $\it {FitCMD}$ were used to decontaminate statistically the CMDs and to perform isochrone fitting, respectively. Importantly, these tools take into account the photometric errors. The derived parameters are: total stellar mass, cluster age, metallicity, foreground reddening and distance modulus.  The clusters CL\,207 and CL\,208 are moderately reddened (with $E(J-K_{S})$ around 0.55 mag), are relatively close to the Sun (with distance around 3 kpc), and have masses less than 2000 $M_{\odot}$ and solar metallicity. The cluster CL\,204 is an exception. Most probably this object is a  dissolving  cluster, being the oldest one in the sample (7.5 Gyr), metal poor ($[Fe/H]$=$-$1.28 dex.), and with a very small total mass of $\approx$ 400M$_\odot$.

\begin{figure*}
\includegraphics[width=15cm]{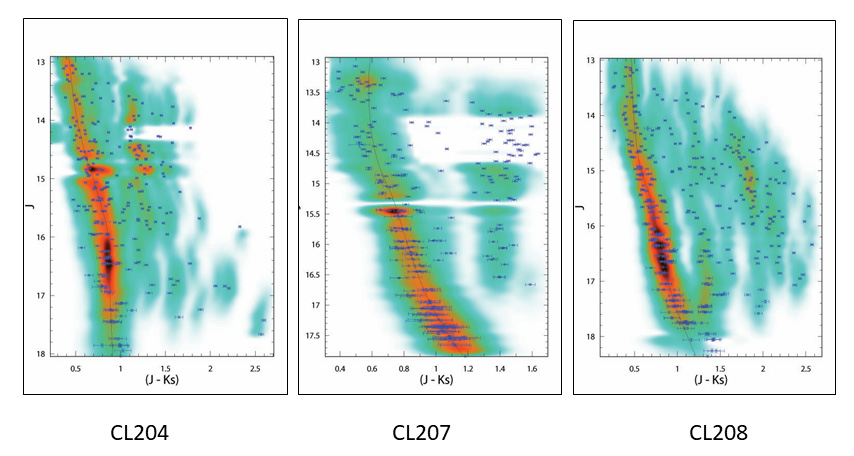}
\caption{Hess color magnitude diagrams of the cluster candidates VVVX CL\,204, CL\,207 and CL\,208.
} 
\label{VVVX_oc_cmd}
\end{figure*}

The Hess color magnitude diagrams of the cluster candidates CL\,204, CL\,207 and CL\,208 are plotted in Fig.\,\ref{VVVX_oc_cmd} and the derived parameters are listed in Table\,\ref{oc_known}.

\section{Known clusters in the  area}~\label{known}

\citet{Bica19} summarized the known open clusters or candidates at that time and listed in total about 10\,000 objects; 2454 of them fall within the area studied here.  From this subset,  817 were classified by the authors as ``embedded cluster (EC)'', ``embedded cluster candidate (ECC)'' or ``embedded group (EGr)''.  Our visual search didn't recover 90 candidates listed in the catalog, mainly from the WISE-based discoveries \citep{Cam16}. On the VVVX images they typically appear as one or two bright stars or windows in the dust screen. The coordinates and designation names as in \citet{Bica19} catalog are listed in Table\,\ref{bica_false}. 

\begin{figure*}
\includegraphics[width=15cm]{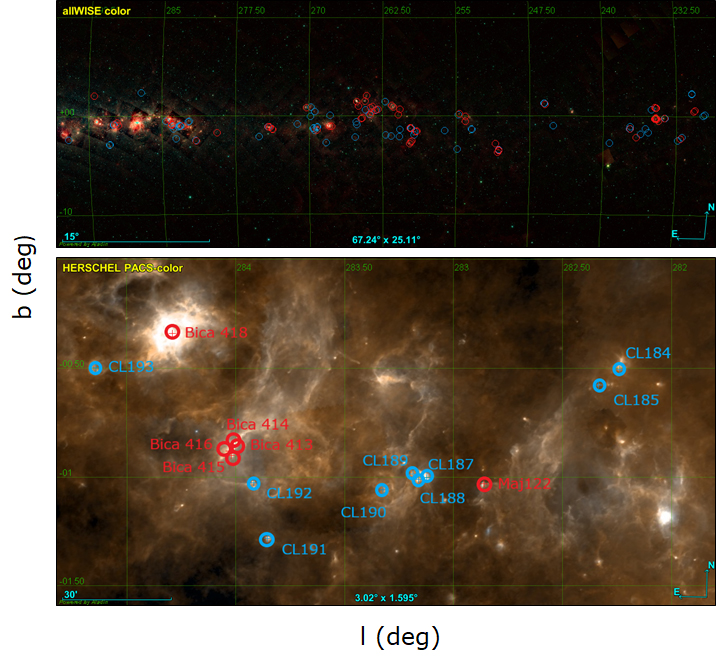}
\caption {The distribution of visually confirmed obscured clusters from \citet{Bica19} (red circles) and newly proposed in this paper VVVX clusters (blue circles) overplotted on \it {allWISE} and {\it Herschel}  images.  } 
\label{bica_vvvx_all}
\end{figure*}

On the other hand, we confirmed by the same method 71 objects,  classified by the authors as embedded clusters and visually connected with  the surrounding nebulosity in the VVVX images. They are verified also on the {\it Herschel} and {\it allWISE} color compositions. 
Figure \ref{bica_vvvx_all} shows the distribution of Bica's objects on the {\it allWISE}, with a zoomed {\it Herschel} area for illustration. Note, that we also overplotted the candidates proposed in this paper (blue circles). The coordinates and designation names as in \citet{Bica19} catalog are listed in Table\,\ref{bica_confirmed}.

\begin{figure*}
\includegraphics[width=15cm]{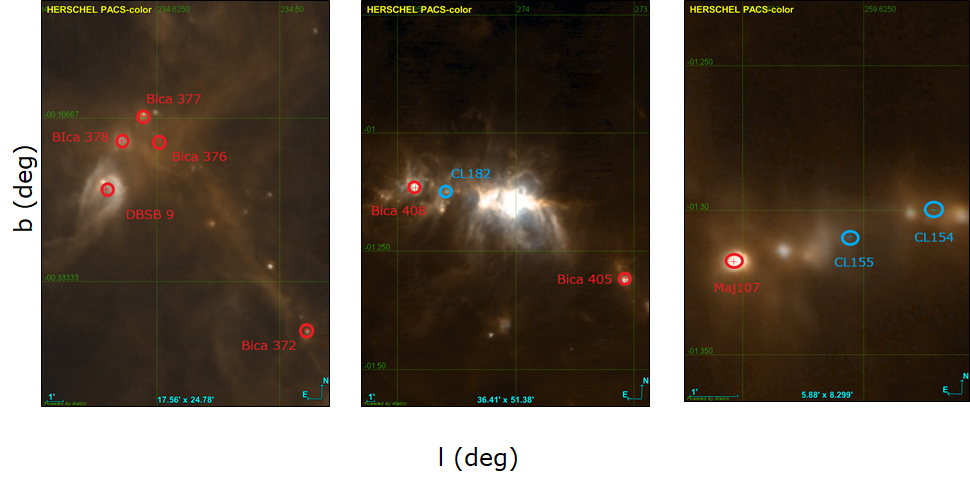}
 \caption{ {\it Herschel} PACS (70\,$\mu$m and 160\,$\mu$m) color composition with overplotted group of clusters. North is up, East is to the left.}
\label{her_area1}
\end{figure*}

We chose 15 of Bica's objects, projected close to our newly discovered candidates or visually on the same nebulosity; with enough resolved member stars and without any determined basic parameters in the literature. Figure \ref{her_area1} shows some examples.

\begin{figure*}
\includegraphics[width=17cm]{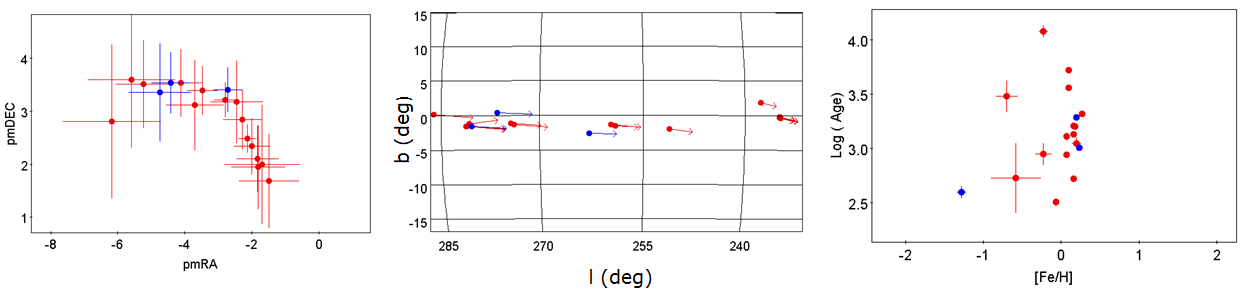}
 \caption{Parameters of the open cluster sample. The blue color stands for the newly proposed candidates.} 
\label{natalia_par}
\end{figure*}

Then we performed color-magnitude analysis, using the same technique as for our open cluster candidates (Sec.\,\ref{open_cl}). For the most probable cluster member stars -- as identified with $\it {CLEAN_{phot}}$ and $\it {FitCMD}$ -- we also report {\it Gaia} Data Release 2 distances from \cite{Bailer18}. For each cluster we adopted an average distance obtained by fitting a Gaussian function to the distribution of the Gaia distance determinations of individual members. We found a good agreement between the astrometric and the $\it {FitCMD}$-derived distances, with measured correlation of 92\% of the linear fit and some systematic effect of 200 pc, where the Gaia distances put the objects closer to the Sun. The derived cluster parameters are listed in Table\,\ref{oc_known} (alongside the new VVVX clusters), and the CMDs are shown in Fig.\,\ref{cmd_nat}. The sample contains not very obscured ($E(J-K)$ between 0.18 and 0.55 mag), low mass (72\% have masses below 2000\,M$_{\odot}$) and relatively young (between 20 and 650 Myr) clusters. Most of the clusters show solar metallicity. The clusters CL\,204 and Teu\,215 are exceptions, showing higher masses and older ages. In Fig.\,\ref{natalia_par} are shown some relations with the above derived basic cluster parameters.  In general, these relations follow the expected distribution for typical open clusters, for example, decreasing $\rm [Fe/H]$  with increasing cluster age.  The {\it Gaia} derived proper motions show that all seem to be moving in the same direction, indicating that they belong to the disk population.

The rest of the \citet{Bica19} clusters, classified by these authors as ``embedded'' and projected on the investigated area, are known and relatively big young clusters, or cannot be associated with any projected nebulosity in the VVVX images. Their analysis is outside the scope of this paper. We also do not include any of the projected in the area of  massive young clusters such as NGC\,3603, Westerlund\,2 and the Carina Nebula Complex.

\section{Spatial distribution of the sample} \label{spacial}

Combining all 60 clusters with the above derived distances  (42 from Table\,\ref{candidates} (regular font) and 18 from Table\,\ref{oc_known})  allows us to look for spatial clustering. Indeed, the visual inspection of the objects identified several groups of clusters. 

The group containing Bica\,376, Bica\,377, Bica\,378, BDSB\,9, Bica\,372 is shown in Fig.\,\ref{her_area1}, left panel. Judging from the network of filaments connecting the dust emission density peaks, as well as similar withing the errors distances, these objects seem to be physically connected and to lie on the same cloud. Another group of clusters is shown in Fig.\,\ref{her}. The mean distance of CL\, 169, 171, 172, 174 and 175 is determined as 2.3$\pm$0.3\,kpc, thus most probably they belong to the same cloud. One more cluster candidate, CL\, 170, is projected in the area, without distance measurements, and we assume the mean distance of the group as the distance to this object. The group Bica\,405 and Bica\,408 shows mean distance of 3.4$\pm0.3$\,kpc. We can adopt this value for the cluster CL\,182.  The group CL\,187, 188, 189 has distance of 3.7 $\pm0.4$ kpc, so this value can be adopted for CL\,190. Maj\,107 has 2.9$\pm0.1$, so we assume that the clusters CL\,154 and 155 are placed on the same distance.  We mark the adopted distances in Table\,\ref{candidates} with bold font.

\section{Summary} 
In the latest observed area of the VVVX survey we report a catalog of 88 new cluster candidates. Radial velocities,
available in the literature for 42 of them allow us to determine their kinematic heliocentric distances, placing most of them between 2 and 4 kpc. These are small clusters, with linear diameters between  0.2 and 1.4 pc.  Comparison with fluxes of the emission clumps in the dust continuum of {\it Herschel} and {\it allWISE} shows that 80 \% of them most probably are still associated with their cold dust.   

Additionally, nine candidates are proposed to be older open cluster candidates. Three of them have sufficient number of well-resolved stellar members, which allow us to determine some basic cluster parameters. We confirm their nature as older, low-mass open clusters. VVVX CL\,204 could be ``dissolving'' cluster, according to its low total cluster mass, low metallicity and older age. 

During our visual inspection of VVVX image we did not recover  90 embedded cluster candidates listed in \citet{Bica19}, and confirmed 71 of them on the base of their appearance on the high quality VVVX images. Photometric analysis of 15 clusters from \citet{Bica19} catalog shows that they have ages above 20\,Myr, masses below 2000\,M$_{\odot}$ and in general follow the motion of the disk, according to their proper motion distribution. 

The distance determinations allow us to outline some groups of clusters, most probably formed within the same dust complex. When possible, the mean distance value is assigned to nearby projected clusters, without their own measurements. In general, our candidates follow the network of filaments structures of the renaming dust.  

Thus, in the this VVVX area, we found recent star formation, which produce small size and young clusters, in addition to the well know, massive young clusters of the MW, including NGC\,3603, Westerlund\,2 and the Carina Nebula Complex.

\section*{Acknowledgements}

We gratefully acknowledge data from the ESO Public Survey program ID 198.B-2004 taken with the VISTA telescope, and products from the Cambridge Astronomical Survey Unit (CASU). This work has made use of data from the European Space Agency (ESA) mission {\it Gaia}\footnote{\url{https://www.cosmos.esa.int/gaia}}, processed by the {\it Gaia} Data Processing and Analysis Consortium (DPAC)\footnote{\url{https://www.cosmos.esa.int/web/gaia/dpac/consortium}}. Funding for the DPAC has been provided by national institutions, in particular the institutions participating in the {\it Gaia} Multilateral Agreement. Support is provided by the ANID, Millennium Science Initiative, PROG. ICM ANID, ICN12-009, awarded to the Millennium Institute of Astrophysics (MAS).  J.A.-G. also acknowledges support from Fondecyt Regular 1201490. S.R.A. acknowledges support from the FONDECYT Iniciaci\'on project 11171025, the FONDECYT Regular project 1201490, and the CONICYT + PAI ``Concurso Nacional Inserci\'on de Capital Humano Avanzado en la Academia 2017'' project PAI 79170089.  This research has made use of the SIMBAD database, operated at CDS, Strasbourg, France. We are grateful to the referee Dr. Joana Ascenso for very useful comments, which greatly improve the manuscript.

Data availability: The data underlying this article are available in the article and in its online supplementary material.





\newpage

\bibliographystyle{mnras}


\appendix

\section{The catalog of newly discovered VVVX Cluster Candidates.} In Table\,\ref{candidates} are tabulated the newly proposed candidate star clusters. The first column gives the identification, followed by the equatorial coordinates of the center, eye-ball measured apparent cluster radius in arcmin, the name of the corresponding VVVX tile, distance to the candidate in kpc, {\it Herschel} SPIRE 500$\mu$m and WISE W4 (22$\mu$m) flux in Jy, calculated mass of the dust in solar masses and any known nearby objects such as IR sources,YSOs, \ion{H}{ii} regions, dense cores (when available), taken from the SIMBAD database (\url{http://simbad.u-strasbg.fr/simbad/}).

\begin{table*}
\caption{Catalog of newly proposed VVVX clusters.}
\begin{small}\tabcolsep=5pt
\begin{tabular}{lccllcclm{7cm}} 
\hline
\multicolumn{1}{l}{Name}&	\multicolumn{1}{c}{$\alpha\delta(J2000)$}&\multicolumn{1}{c}{Radius}&	\multicolumn{1}{c}{VVVX}&	\multicolumn{1}{c}{Kinem. Dist.} &	\multicolumn{1}{c}{F500\,$\mu$m} &\multicolumn{1}{c}{F22\,$\mu$m} & \multicolumn{1}{c}{M$_{\rm dust}$}   &	\multicolumn{1}{c}{Comments}	\\
\multicolumn{1}{c}{}&	\multicolumn{1}{c}{$^\circ$} & \multicolumn{1}{c}{arcmin} & \multicolumn{1}{c}{tile}& \multicolumn{1}{c}{kpc} &  \multicolumn{1}{c}{Jy} & \multicolumn{1}{c}{Jy}  & \multicolumn{1}{c}{M$_\odot$}& \\
\hline
VVVX CL121  &	109.7116 $-$18.3699	&	0.42	&	e686	&				        &				&283 &      &IRAS 07166-1816 -- Far-IR source	\\	
VVVX CL122	&	109.7168 $-$18.3805	&	0.42	&	e686	&				        &				&    &      &no known sources	\\	
VVVX CL123	&	109.7930 $-$17.8238	&	0.25	&	e685	&				        &				&358 &      &MSX6C G231.9005-02.1378 -- HII (ionized) region	\\	
VVVX CL124	&	110.7560 $-$14.6912	&	0.60	&	e1091	&	3.2$\pm$1.0	  & 264   &387 &   386&[BNM96] 229.570+0.150 -- HII (ionized) region 	\\	
VVVX CL125	&	110.7717 $-$14.9151	&	0.83	&	e1091	&				        &				&163 &      &no known sources	\\	
VVVX CL126	&	110.7734 $-$14.7108	&	0.92	&	e1091	&				        &				&173 &      &no known sources	\\	
VVVX CL127	&	111.9843 $-$20.6689	&	0.47	&	e1005	&	7.4$\pm$1.3	  &	171	  &269 &1347	&IRAS 07257-2033 -- Far-IR source 	\\	
VVVX CL128	&	112.2242 $-$21.4514	&	0.10	&	e1005 &	2.3$\pm$0.6	  &	253 	&190 &183	  &IRAS 07267-2120 -- Far-IR source 	\\	
VVVX CL129	&	112.4938 $-$18.4728	&	1.48	&	e1048	&				        &				&		 &	    &no known sources	\\	
VVVX CL130	&	112.5211 $-$20.7384	&	0.65	&	e1005 &	6.7$\pm$1.1	  &	215   &279 &1360	&BRAN 33A -- Star forming region 	\\	
VVVX CL131	&	112.9961 $-$24.6527	&	0.42	&	e690	&				        &				&211 &	    &PN G239.3-02.7 -- Possible Planetary Nebula 	\\	
VVVX CL132	&	113.0043 $-$24.6423	&	0.33	&	e690	&				        &				&170 &	    &no known sources	\\	
VVVX CL133	&	113.3238 $-$14.8495	&	0.42	&	e768	&				        &				&184 &	    &no known sources	\\	
VVVX CL134	&	113.3329 $-$22.1824	&	0.43	&	e1006	&				        &				&471 &	    &BRAN 45 -- Interstellar matter	\\	
VVVX CL135	&	113.3457 $-$14.8520	&	0.58	&	e768	&				        &				&169 &	    &no known sources	\\	
VVVX CL136	&	113.3731 $-$14.8568	&	0.67	&	e768	&				        &				&181 &	    &no known sources	\\	
VVVX CL137	&	113.4895 $-$22.0031	&	0.43	&	e1006 &	1.7$\pm$0.7	  &	200 	&194 &	82	&BRAN 47 -- HII (ionized) region 	\\	
VVVX CL138	&	113.9709 $-$19.2728	&	0.97	&	e1094	&	3.5$\pm$0.8	  &		 		&190 &	    &IRAS 07336-1909 -- HII (ionized) region 	\\	
VVVX CL139	&	117.8957 $-$28.7996	&	0.35	&	e1056	&	7.9$\pm$1.4	  &	141 	&    &	1244&IRAS 07495-2840 -- Far-IR source 	\\	
VVVX CL140	&	118.5640 $-$35.0801	&	0.33	&	e615	&				        &				&163 &	    &no known sources	\\	
VVVX CL141	&	118.5682 $-$35.0694	&	0.27	&	e615	&				        &				&170 &	    &IRAS 07523-3456 -- Far-IR source	\\	
VVVX CL142	&	118.7334 $-$34.8285	&	0.58	&	e615	&				        &				&340 &	    &VLA G250.6504-03.4733 -- HII (ionized) region	\\	
VVVX CL143	&	120.6256 $-$28.3981	&	0.40	&	e1147	&	4.8$\pm$0.9	  &		 		&200 &	    &BRAN 98 -- Interstellar matter 	\\	
VVVX CL144	&	120.6895 $-$28.4340	&	0.37	&	e1147	&				        &				&345 &	    &G246.0702+01.2909 1 -- YSO Cand.	\\	
VVVX CL145	&	121.5830 $-$38.4280	&	0.50	&	e618	&	9.6$\pm$1.5	  &		 		&214 &	    &IRAS 08045-3816 -- Far-IR source 	\\	
VVVX CL146	&	122.6767 $-$36.0521	&	0.78	&	e1017 &	1.1$\pm$0.7	  &	293 	&432 &	54	&IRAS 08088-3554 -- HII (ionized) region 	\\	
VVVX CL147	&	122.7522 $-$36.0806	&	0.75	&	e1017 &				        &				&371 &	    &2MASX J08110108-3605007 -- IR source	\\	
VVVX CL148	&	124.4691 $-$38.4564	&	0.75	&	e1019 &	8.4$\pm$1.3	  &	153	  &186 &	1534&IRAS 08160-3818 -- Far-IR source 	\\	
VVVX CL149	&	125.1917 $-$36.3529	&	0.32	&	e1108	&				        &				&248 &	    &IRAS 08188-3611 -- Far-IR source	\\	
VVVX CL150	&	125.7669 $-$41.9324	&	0.20	&	e704	&				        &				&438 &	    &2MASX J08230395-4155551 -- YSO Candidate	\\	
VVVX CL151	&	126.3285 $-$41.2428	&	0.13	&	e1021 &	5.6$\pm$0.9	  &	126   &221 &	572	&IRAS 08235-4104 -- Far-IR source 	\\	
VVVX CL152	&	126.5715 $-$40.8113	&	0.23	&	e1021 &	1.8$\pm$0.7	  &	203   &440 &	98	&[MHL2007] G259.0453-01.5559 1 -- YSO Cand. 	\\	
VVVX CL153	&	126.6394 $-$41.0106	&	0.25	&	e1021 &	2.0$\pm$1.3	  &	162   &258 &	93	&IRAS 08247-4050 -- Far-IR source 	\\	
VVVX CL154	&	127.2737 $-$41.1136	&	0.77	&	e1021 &{\bf2.9$\pm$0.1}&		  &181 &	    &no known sources	\\	
VVVX CL155	&	127.2863 $-$41.1424	&	0.47	&	e1021 &{\bf2.9$\pm$0.1}&		  &    &	    &no known sources	\\	
VVVX CL156	&	127.3320 $-$42.5772	&	0.27	&	e1022 &				        &				&191 &	    &no known sources	\\	
VVVX CL157	&	128.0324 $-$43.2308	&	0.93	&	e1023 &	2.2$\pm$0.8	  &	217   &478 &	149	&IRAS 08303-4303 -- HII (ionized) region 	\\	
VVVX CL158	&	128.0538 $-$42.0740	&	0.35	&	e1022 &	2.1$\pm$0.8	  &	150   &205 &	97	&IRAS 08304-4153 -- Far-IR source 	\\	
VVVX CL159	&	128.6130 $-$43.5796	&	0.45	&	e1023	&	1.9$\pm$0.9	  &	172   &189 &	85	&IRAS 08327-4324 -- Far-IR source 	\\	
VVVX CL160	&	131.0680 $-$46.1999	&	0.33	&	e708	&	5.1$\pm$0.9	  &	158   &286 &	595	&IRAS 08426-4601 -- HII (ionized) region 	\\	
VVVX CL161	&	132.1944 $-$42.7424	&	0.25	&	e1114	&				        &				&    &   		&no known sources	\\	
VVVX CL162	&	132.1958 $-$42.7510	&	0.17	&	e1114	&				        &				&    &		  &no known sources	\\	
VVVX CL163	&	132.2759 $-$45.2612	&	0.42	&	e1070	&	3.6$\pm$0.8	  &	163   &318 &	309	&IRAS 08473-4504 -- HII (ionized) region 	\\
VVVX CL164  &	132.3000 $-$41.5764	&	0.45	&	e1158	&	1.0$\pm$1.4  	&	 		  &230 &	 	  &ESO-HA 196 -- Emission-line Star 	\\	
VVVX CL165	&	132.3061 $-$43.6065	&	0.48	&	e1114	&	9.7$\pm$2.8	  &	202   &203 &	2724&2MASS J08491394-4336092 -- YSO 	\\	
VVVX CL166	&	132.6621 $-$45.1388	&	0.42	&	e1070	&	1.6$\pm$0.8	  &	185   &	   &	69	&TGU H1699 -- Dark Cloud (nebula) 	\\	
VVVX CL167	&	133.1393 $-$48.8477	&	0.25	&	e710	&				        &				&178 &	    &IRAS 08509-4839 -- Far-IR source	\\	
VVVX CL168	&	133.1728 $-$43.6181	&	0.60	&	e1114	&	1.4$\pm$0.8	  &	 	 		&	 	 &	    &IRAS 08509-4325 -- Dense core 	\\	
VVVX CL169	&	135.8099 $-$48.9216	&	0.85	&	e1028	&	2.2$\pm$0.9	  &	243   &503 &	172	&IRAS 09015-4843 -- HII (ionized) region 	\\	
VVVX CL170	&	135.8627 $-$48.5216	&	0.60	&	e1028	&	{\bf2.3$\pm$0.3}&		  &351 &	    &IRAS 09017-4819 -- IR source	\\	
VVVX CL171	&	135.8838 $-$48.4790	&	0.47	&	e1073	&	2.5$\pm$1.0	  &	395   &1019&	358	&IRAS 09018-4816 -- HII (ionized) region 	\\	
VVVX CL172	&	135.8840 $-$48.4616	&	0.65	&	e1073	&	2.7$\pm$0.8	  &	395   &664 &	411	&[HBM2005] G269.15-1.13 -- mm radiosource 	\\	
VVVX CL173	&	135.8992 $-$46.7214	&	0.28	&	e1117	&	2.2$\pm$0.9	  &	240   &171 &	166	&IRAS 09018-4631 -- HII (ionized) region 	\\	
VVVX CL174	&	136.0951 $-$48.9045	&	0.68	&	e1028	&	2.2$\pm$0.8	  &	249   &349 &	169	&IRAS 09026-4842 -- Far-IR source 	\\	
VVVX CL175	&	136.1257 $-$48.8259	&	0.62	&	e1028	&	2.0$\pm$0.9	  &	254   &342 &	150	&IRAS 09028-4837 -- HII (ionized) region 	\\	
VVVX CL176	&	137.7850 $-$48.2656	&	0.42	&	e1073	&				        &				&488 &      &[BNM96] 269.854-0.063 -- HII (ionized) reg. 	\\	
VVVX CL177	&	137.8902 $-$47.6336	&	1.35	&	e1118	&	1.9$\pm$0.8	  &		 		&220 &      &IRAS 09098-4725 -- Far-IR source  	\\	
VVVX CL178	&	138.3333 $-$49.7635	&	0.60	&	e1074	&				        &				&172 &      &IRAS 09116-4933 -- Far-IR source	\\	
VVVX CL179	&	138.7102 $-$47.5774	&	0.82	&	e1118	&				        &				&279 &      &no known sources	\\	
VVVX CL180	&	138.7209 $-$47.6009	&	0.60	&	e1118	&	2.2$\pm$0.9	  &	 	 	 	&441 &      &IRAS 09131-4723 -- Far-IR source 	\\	
\hline
\end{tabular}
\label{candidates}
\end{small}
\end{table*}

\begin{table*}
\contcaption{Catalog of newly proposed VVVX clusters.}
\begin{small}\tabcolsep=5pt
\begin{tabular}{lccllcclm{7cm}} 
\hline
\multicolumn{1}{l}{Name}&	\multicolumn{1}{c}{$\alpha\delta(J2000)$}&\multicolumn{1}{c}{Radius}&	\multicolumn{1}{c}{VVVX}&	\multicolumn{1}{c}{Kinem. Dist.} &	\multicolumn{1}{c}{F500\,$\mu$m} &\multicolumn{1}{c}{F22\,$\mu$m} & \multicolumn{1}{c}{M$_{\rm dust}$}   &	\multicolumn{1}{c}{Comments}	\\
\multicolumn{1}{c}{}&	\multicolumn{1}{c}{$^\circ$} & \multicolumn{1}{c}{arcmin} & \multicolumn{1}{c}{tile}& \multicolumn{1}{c}{kpc} &  \multicolumn{1}{c}{Jy} & \multicolumn{1}{c}{Jy}  & \multicolumn{1}{c}{M$_\odot$}& \\
\hline
VVVX CL181  &	141.2859 $-$53.4727	&	0.28	&	e1032	&	7.7$\pm$1.0	        &135   &226	&1135	  &IRAS 09235-5315 -- Far-IR source 	\\	
VVVX CL182	&	141.2958 $-$52.0718	&	0.18	&	e1031	&	{\bf3.4$\pm$0.3}	  &      &261	&		    &no known sources	\\	
VVVX CL183	&	149.5112 $-$57.9644	&	0.67	&	e719	&	3.1$\pm$0.7	        &	 	 	 &464	&	      &IRAS 09563-5743 -- HII (ionized) region 	\\	
VVVX CL184	&	152.6955 $-$56.7787	&	0.38	&	e1082	&	3.7$\pm$0.9	        &277   &227	&536	  &[BYF2011] 23a -- Dense core 	\\	
VVVX CL185	&	152.7636 $-$56.8935	&	0.38	&	e1082	&				              &			 &183 &		    &IRAS 10092-5638 -- Far-IR source	\\	
VVVX CL186	&	152.8120 $-$58.9722	&	0.67	&	e721	&	3.9$\pm$0.6	        &	 	 	 &373 &	    	&[BYF2011] 38 -- Molecular Cloud 	\\	
VVVX CL187	&	153.5177 $-$57.6919	&	0.90	&	e1082	&	3.8$\pm$0.7	        &	271  &549	&560	  &[BYF2011] 36c -- Dense core 	\\	
VVVX CL188	&	153.5682 $-$57.7203	&	0.50	&	e1082	&	3.4$\pm$0.9	        &	239  &329	&382	  &[BYF2011] 36d -- Dense core 	\\	
VVVX CL189	&	153.6358 $-$57.7125	&	0.28	&	e1082	&	3.8$\pm$0.7        	&	245  &274	&493	  &[BYF2011] 36a -- Dense core 	\\	
VVVX CL190	&	153.7691 $-$57.8498	&	0.38	&	e1082	&	{\bf3.7$\pm$0.4}    &		   &420	&       &GAL 283.33-01.05 -- HII (ionized) region	\\	
VVVX CL191	&	154.3684 $-$58.3307	&	0.50	&	e1038	&			                &			 &317 &	      &IRAS 10156-5804 -- Far-IR source 	\\	
VVVX CL192	&	154.7375 $-$58.1542	&	0.40	&	e1084	&			                &			 &312 &		    &AGAL G283.916-01.024 -- sub-mm source	\\	
VVVX CL193	&	156.4277 $-$58.0976	&	0.73	&	e1083	&				              &			 &  	&       &AGAL G284.631-00.511 -- sub-mm source	\\	
VVVX CL194	&	163.0621 $-$56.7763	&	0.33	&	e806	&	3.7$\pm$0.9	        &	 	 	 &  	&       &IRAS 10501-5630 -- HII (ionized) region 	\\	
VVVX CL195	&	164.3552 $-$62.8852	&	0.33	&	e725	&	3.3$\pm$4.4       	&	 	 	 &308	&       &IRAS 10554-6237 -- possible Herbig Ae/Be 	\\	
VVVX CL196	&	164.3915 $-$62.9814	&	0.30	&	e725	&	4.3$\pm$0.9	        &	 	 	 &447	&   	  &IRAS 10555-6242 -- HII (ionized) region 	\\	
VVVX CL197	&	165.2948 $-$60.9496	&	0.48	&	e1091	&	8.8$\pm$0.8	        &	325  &578	&3575   &IRAS 10591-6040 -- Composite object 	\\	
VVVX CL198	&	165.4324 $-$60.9586	&	0.80	&	e1091	&		 		              &		 	 &462	&       &IRAS 10597-6041 -- Far-IR source	\\	
VVVX CL199	&	173.6373 $-$63.9679	&	0.33	&	e728	&				              &			 &335	&       &IRAS 11322-6341 -- YSO Candidate	\\
VVVX CL200	&	109.8646 $-$17.8712	&	0.42	&	e685	&                     &      &    &       &no known sources          \\
VVVX CL201	&	111.8978 $-$23.0899	&	1.00	&	e689	&                     &      &    &       &no known sources       \\
VVVX CL202	&	121.6348 $-$37.5006	&	0.55	&	e700	&                     &      &    &       &no known sources       \\
VVVX CL203	&	124.1511 $-$36.2379	&	0.30	&	e1062	&                     &      &    &       &no known sources        \\
VVVX CL204	&	128.4525 $-$44.4484	&	0.75	&	e706	&                     &      &    &       &no known sources          \\
VVVX CL205	&	134.4359 $-$43.1951	&	0.58	&	e1160 &                     &      &	  &       &no known sources           \\
VVVX CL206	&	137.4686 $-$48.8665	&	0.83	&	e1073	&                     &      &    &       &GSC 08173-00182 -- OB+?         \\
VVVX CL207	&	145.6299 $-$52.4305	&	1.25	&	e1123 &                     &      &    &       &no known sources                       \\
VVVX CL208	&	148.5833 $-$56.4236	&	1.00	&	e1035 &                     &      &    &       &no known sources                        \\
\hline
\end{tabular}
\label{candidates: continued}
\end{small}
\end{table*}


\section{Known clusters in the region.}  

\begin{table*}
\caption{False positive objects from \citet{Bica19} catalog as revealed by VVVX. The first column gives the sequence number in this table, followed by the identification and galactic coordinates of the center taken from Bica et al.}
\begin{small}\tabcolsep=4.7pt
\begin{tabular}{clcc|clcc|clcc} 
\hline
\multicolumn{1}{c}{Seq.} & \multicolumn{1}{l}{Name} & \multicolumn{1}{c}{GLON}	& \multicolumn{1}{c|}{GLAT} & 
\multicolumn{1}{c}{Seg.} & \multicolumn{1}{l}{Name}	& \multicolumn{1}{c}{GLON}	& \multicolumn{1}{c|}{GLAT} & 
\multicolumn{1}{c}{Seq.} & \multicolumn{1}{l}{Name}	& \multicolumn{1}{c}{GLON}	& \multicolumn{1}{c}{GLAT}   \\
\multicolumn{1}{c}{Numb.} & \multicolumn{1}{c}{} & \multicolumn{1}{c}{deg} & \multicolumn{1}{c|}{deg} &
\multicolumn{1}{c}{Numb.} & \multicolumn{1}{c}{} & \multicolumn{1}{c}{deg} & \multicolumn{1}{c|}{deg} & 
\multicolumn{1}{c}{Numb.} & \multicolumn{1}{c}{} & \multicolumn{1}{c}{deg} & \multicolumn{1}{c}{deg} \\
\hline		
1	  &	Bica 370	&	234.32	&	-0.45	&	31	&	Cmg 989	  &	236.98	&	-2.69	&	61	&	Cmg 378	&	279.12	&	-1.22	\\
2	  &	Cmg 264	  &	264.54	&	0.61	&	32	&	Bica 410	&	283.11	&	-0.97	&	62	&	Cmg 379	&	279.12	&	-1.22	\\
3	  &	Cmg 266	  &	264.97	&	0.26	&	33	&	Bica 411	&	283.43	&	-0.99	&	63	&	Cmg 381	&	279.42	&	-0.96	\\
4	  &	Cmg 339	  &	275.72	&	-2.23	&	34	&	Bica 412	&	283.58	&	-0.97	&	64	&	Cmg 382	&	279.43	&	-1.69	\\
5	  &	Cmg 340	  &	275.80	&	-2.2	&	35	&	Cmg 1032	&	239.65	&	-2.08	&	65	&	Cmg 383	&	279.5	&	-1.52	\\
6	  &	Cmg 341	  &	275.88	&	-2.15	&	36	&	Cmg 1070	&	249.96	&	-3.49	&	66	&	Cmg 384	&	279.52	&	-1.73	\\
7	  &	Cmg 342	  &	275.97	&	-2.03	&	37	&	Cmg 1073	&	250.13	&	-2.97	&	67	&	Cmg 385	&	279.57	&	-1.42	\\
8	  &	Bica 393	&	262.26	&	1.45	&	38	&	Cmg 1086	&	252.08	&	-4.24	&	68	&	Cmg 386	&	279.67	&	-1.02	\\
9	  &	Cmg 240	&	260.81	&	0.19	&	39	&	Cmg 221	&	259.09	&	-1.78	&	69	&	Cmg 387	&	279.93	&	-0.4	\\
10	&	Cmg 244	&	261.04	&	1.03	&	40	&	Cmg 225	&	259.31	&	-1.62	&	70	&	Cmg 388	&	280.00	&	-1.2	\\
11	&	Cmg 260	&	262.45	&	-3.06	&	41	&	Cmg 228	&	259.57	&	-1.44	&	71	&	Cmg 389	&	280.02	&	-1.42	\\
12	&	Cmg 269	&	265.82	&	0.92	&	42	&	Cmg 229	&	259.64	&	-1.31	&	72	&	Cmg 390	&	280.37	&	-1.68	\\
13	&	Cmg 270	&	265.84	&	1.1	&	43	&	Cmg 232	&	259.77	&	-2.78	&	73	&	Cmg 392	&	280.43	&	-1.74	\\
14	&	Maj 106	&	259.02	&	-1.91	&	44	&	Cmg 236	&	259.91	&	-1.33	&	74	&	Reyle-Robin 1	&	269.27	&	-1.53	\\
15	&	Cmg 1070	&	250.01	&	-3.34	&	45	&	Cmg 237	&	260.08	&	-1.38	&	75	&	BH 23	&	254.09	&	-0.97	\\
16	&	Cmg 222	&	259.12	&	-1.74	&	46	&	Cmg 242	&	260.95	&	0.87	&	76	&	Bica 380	&	234.84	&	-0.12	\\
17	&	Cmg 1087	&	252.58	&	-4.01	&	47	&	Cmg 257	&	261.78	&	-3.04	&	77	&	Cmg 334	&	275.14	&	-1.07	\\
18	&	Cmg 268	&	265.68	&	1.07	&	48	&	Cmg 267	&	265.51	&	1.34	&	78	&	Cmg 253	&	261.56	&	-2.51	\\
19	&	Alessi 43	&	262.64	&	1.38	&	49	&	Cmg 289	&	268.62	&	0.45	&	79	&	Cmg 417	&	281.56	&	-2.48	\\
20	&	Cmg 1038	&	240.75	&	-1.62	&	50	&	Cmg 290	&	268.65	&	0.18	&	80	&	Cmg 420	&	281.65	&	-2.07	\\
21	&	Cmg 377	&	279.04	&	-1.13	&	51	&	Cmg 292	&	269.61	&	0.94	&	81	&	Cmg 421	&	281.69	&	-2.06	\\
22	&	Cmg 393	&	280.44	&	-1.83	&	52	&	Cmg 293	&	269.67	&	1.07	&	82	&	Cmg 1075	&	250.57	&	-3.43	\\
23	&	Cmg 988	&	236.79	&	-2.25	&	53	&	Cmg 295	&	269.87	&	0.78	&	83	&	Cmg 336	&	275.57	&	-2.2	\\
24	&	Cmg 231	&	259.73	&	-1.32	&	54	&	Cmg 297	&	269.97	&	0.84	&	84	&	Cmg 419	&	281.64	&	-2.54	\\
25	&	Cmg 252	&	261.54	&	-2.32	&	55	&	Cmg 298	&	270.04	&	0.76	&	85	&	Cmg 380	&	279.4	&	-1.68	\\
26	&	Cmg 1017	&	238.88	&	-2.13	&	56	&	Cmg 300	&	270.25	&	-1.26	&	86	&	DBSB 29	&	264.99	&	0.99	\\
27	&	Cmg 993	&	237.22	&	-2.9	&	57	&	Cmg 315	&	270.88	&	-0.72	&	87	&	Bica 396	&	263.5	&	0.93	\\
28	&	Cmg 992	&	237.21	&	-2.52	&	58	&	Cmg 318	&	270.96	&	-1.11	&	88	&	MLG 11	&	263.74	&	0.82	\\
29	&	Cmg 1081	&	250.98	&	-3.1	&	59	&	Cmg 332	&	275.07	&	-1.09	&	89	&	Cmg 255	&	261.59	&	-2.47	\\
30	&	Cmg 985	  &	236.59	&	-2.3	&	60	&	Cmg 359	&	277.23	&	0.64	&	90	&	HD 77343 Group	&	265.46	&	1.46	\\
\hline
\end{tabular}
\label{bica_false}
\end{small}
\end{table*}

\begin{table*}
\caption{Visually confirmed 71 objects from \citet{Bica19} catalog as revealed by VVVX. The first column gives the sequence number in this table, followed by the identification, equatorial  coordinates of the center and assigned classification taken from Bica et al.; the name of the corresponding VVVX tile and the {\it Herschel} SPIRE 500 $\mu$m flux in Jy.}
\begin{small}\tabcolsep=4pt
\begin{tabular}{clccclc|cclcccc} 
\hline
\multicolumn{1}{c}{Seq.} & \multicolumn{1}{l}{Name} &	\multicolumn{1}{c}{RA}	&\multicolumn{1}{c}{DEC} & \multicolumn{1}{l}{Class} 
& \multicolumn{1}{l}{Tile} & 	\multicolumn{1}{c|}{Flux 500\,$\mu$m} & \multicolumn{1}{c}{Seq.} & \multicolumn{1}{l}{Name} &	\multicolumn{1}{c}{RA}	&\multicolumn{1}{c}{DEC} & \multicolumn{1}{l}{Class} & \multicolumn{1}{l}{Tile} & \multicolumn{1}{c}{Flux 500\,$\mu$m} \\

\multicolumn{1}{c}{Numb.}  & \multicolumn{1}{c}{}  & \multicolumn{1}{c}{deg}	&	\multicolumn{1}{c}{deg}  & \multicolumn{1}{c}{}  
& \multicolumn{1}{c}{}& \multicolumn{1}{c|}{Jy} & \multicolumn{1}{c}{Numb.}  & \multicolumn{1}{c}{}  & \multicolumn{1}{c}{deg}	&	\multicolumn{1}{c}{deg}  & \multicolumn{1}{c}{}  & \multicolumn{1}{c}{}  & \multicolumn{1}{c}{Jy} \\
\hline		 
1	&	Bica 366	&	109.7484	&	-18.0493	&	EC	&	e1091	&	260.56$\pm$0.56	&	37	&	Bica 391	&	131.0482	&	-41.2728	&	EC	&	e1114	&				\\
2	&	Bica 356	&	111.6487	&	-15.5345	&	EC	&	e1091	&	 		 	&	38	&	Bica 392	&	131.3454	&	-41.2487	&	EC	&	e1158	&				\\
3	&	Bica 355	&	111.6723	&	-15.4085	&	EC	&	e1091	&	 		 	&	39	&	Bica 394	&	132.1325	&	-42.5997	&	EC	&	e1114	&				\\
4	&	Cmg 986	&	112.1574	&	-21.9610	&	EC	&	e688	&	264.43$\pm$0.13	&	41	&	MLG 10 	&	132.1936	&	-43.5426	&	EC	&	e1114	&	288.29$\pm$0.51	\\
5	&	Mayer 3 	&	112.5197	&	-18.5308	&	EC	&	e1048	&	 		 	&	41	&	DBSB 23 	&	132.1952	&	-42.9101	&	EC	&	e1159	&				\\
6	&	Bica 371	&	112.5686	&	-19.1779	&	EC	&	e1049	&	277.53$\pm$0.35	&	42	&	MLG 8 	&	132.3667	&	-43.2854	&	EC	&	e1114	&	268.90$\pm$0.61	\\
7	&	Bica 373	&	112.7852	&	-19.2598	&	EC	&	e1049	&	276.19$\pm$0.53	&	43	&	MLG 17	&	132.5622	&	-44.4364	&	EC	&	e1070	&	338.27$\pm$0.09	\\
8	&	Bica 374	&	112.8622	&	-19.2883	&	EC	&	e1049	&	313.84$\pm$0.46	&	44	&	MLG 18 	&	132.5920	&	-44.5096	&	EC	&	e1070	&	267.77$\pm$0.37	\\
9	&	Bica 375	&	112.9188	&	-19.2817	&	EC	&	e1049	&	318.16$\pm$0.22	&	45	&	Bica 396	&	132.8547	&	-42.8372	&	EC	&	e1114	&				\\
10	&	DBSB 9	&	112.9462	&	-19.3702	&	EC	&	e1049	&	295.70$\pm$0.14	&	46	&	MLG 11 	&	132.9479	&	-43.0922	&	EC	&	e1114	&				\\
11	&	Bochum 6 	&	112.9499	&	-19.4596	&	EC	&	e1049	&	217.26$\pm$0.44	&	47	&	DBSB 28 	&	134.1138	&	-43.0951	&	EC	&	e1159	&	266.78$\pm$0.37	\\
12	&	Bica 376	&	112.9620	&	-19.2936	&	EC	&	e1049	&	351.10$\pm$0.18	&	48	&	DBSB 29 	&	134.2371	&	-43.9390	&	EC	&	e1115	&	187.61$\pm$0.33	\\
13	&	Bica 378	&	112.9824	&	-19.3286	&	EC	&	e1049	&	322.88$\pm$0.20	&	49	&	DB 94 	&	134.5235	&	-47.3651	&	EC	&	e1027	&	238.40$\pm$0.23	\\
14	&	Bica 377	&	113.0000	&	-19.2966	&	EC	&	e1049	&	315.08$\pm$0.39	&	50	&	MLG 14 	&	134.5506	&	-42.6234	&	EC	&	e1159	&	316.05$\pm$0.38	\\
15	&	Maj 91	&	113.3036	&	-22.1316	&	EC	&	e1006	&	243.41$\pm$0.09	&	51	&	RCW 38 	&	134.7503	&	-47.4968	&	EC	&	e1027	&	395.68$\pm$0.07	\\
16	&	DBSB 7 	&	113.8947	&	-18.7598	&	EC	&	e1094	&	 		 	&	52	&	RCW 3 	&	134.8560	&	-43.7523	&	EC	&	e1160	&	338.27$\pm$0.09	\\
17	&	DBSB 8 	&	113.9163	&	-18.8170	&	EC	&	e1094	&	 		 	&	53	&	Gum 25	&	135.5417	&	-48.7083	&	EC	&	e1028	&	276.98$\pm$0.06	\\
18	&	Bica 379	&	114.0365	&	-18.8567	&	EC	&	e1094	&	 		 	&	54	&	DBSB 36 	&	139.1830	&	-47.9367	&	EC	&	e1118	&				\\
19	&	Cmg 1077	&	118.5627	&	-35.0735	&	EGr	&	e615	&	 		 	&	55	&	Bica 405	&	140.5117	&	-51.9436	&	EC	&	e1031	&	202.82$\pm$0.44	\\
20	&	YEP 1	&	118.6574	&	-34.9108	&	EC	&	e615	&	 		 	&	56	&	Bica 406	&	140.6493	&	-51.9432	&	EC	&	e1031	&	223.25$\pm$0.38	\\
21	&	Cmg 1076	&	118.6574	&	-34.9108	&	EC	&	e615	&	 		 	&	57	&	DBSB 38 	&	141.0975	&	-51.9904	&	EC	&	e1031	&	373.95$\pm$0.28	\\
22	&	Slotegraaf 14	&	118.6951	&	-34.8458	&	EC	&	e615	&	 		 	&	58	&	Bica 407	&	141.1785	&	-52.0252	&	EC	&	e1031	&	357.85$\pm$1.07	\\
23	&	Magakian 303	&	119.6409	&	-34.7959	&	EC	&	e698	&	 		 	&	59	&	Bica 408	&	141.3874	&	-52.1083	&	EC	&	e1031	&	242.74$\pm$0.30	\\
24	&	Bica 385	&	120.4785	&	-28.3832	&	EC	&	e1147	&	 		 	&	60	&	Cmg 380	&	147.3309	&	-55.9953	&	EC	&	e1035	&	202.88$\pm$0.15	\\
25	&	DBSB 16 	&	123.9972	&	-36.1438	&	EC	&	e1062	&	305.60$\pm$0.30	&	61	&	Cmg 415	&	149.3425	&	-57.9856	&	EC	&	e0719	&	193.96$\pm$0.34	\\
26	&	Maj 95 	&	124.4788	&	-35.8808	&	EC	&	e1062	&	284.59$\pm$0.84	&	62	&	Maj 122 	&	153.0845	&	-57.5643	&	EC	&	e1082	&	217.55$\pm$0.49	\\
27	&	Collinder 182 	&	125.2111	&	-36.2147	&	EC	&	e1108	&	177.97$\pm$0.10	&	63	&	Bica 415	&	155.0290	&	-58.1034	&	EC	&	e1083	&				\\
28	&	DBSB 18 	&	125.7159	&	-42.1341	&	EC	&	e704	&	 		 	&	64	&	Bica 413	&	155.0545	&	-58.0644	&	EC	&	e1083	&				\\
29	&	Maj 103 	&	125.7555	&	-41.9311	&	EC	&	e704	&	 		 	&	65	&	Bica 414	&	155.0648	&	-58.0561	&	EC	&	e1083	&				\\
30	&	Maj 104 	&	125.8144	&	-41.7666	&	EC	&	e704	&	 		 	&	66	&	Bica 416	&	155.0656	&	-58.0921	&	EC	&	e1083	&				\\
31	&	Cmg 219	&	126.5274	&	-40.8767	&	EC	&	e1021	&	171.87$\pm$0.05	&	67	&	Bica 418	&	156.0267	&	-57.7777	&	EC	&	e1083	&	269.22$\pm$0.15	\\
32	&	Cmg 218	&	126.5799	&	-40.8117	&	EC	&	e1021	&	211.83$\pm$0.21	&	68	&	Maj 131 	&	164.4179	&	-60.7654	&	EC	&	e1087	&	223.53$\pm$0.09	\\
33	&	Maj 107 	&	127.3060	&	-41.1815	&	EC	&	e1021	&	134.89$\pm$0.09	&	69	&	Maj 134 	&	164.8180	&	-60.5807	&	EC	&	e1087	&	236.91$\pm$0.19	\\
34	&	Cmg 230	&	127.3910	&	-41.1846	&	EC	&	e1021	&	127.68$\pm$0.32	&	70	&	Maj 137 	&	165.2671	&	-60.8460	&	EC	&	e1087	&	317.81$\pm$0.20	\\
35	&	MLG 3 	&	129.8329	&	-41.3375	&	EC	&	e1112	&	233.77$\pm$0.21	&	71	&	DBSB 64	&	171.1679	&	-58.9359	&	EC	&	e1178	&	 		 	\\
36	&	MLG 1 	&	130.2729	&	-40.8669	&	EC	&	e1112	&				&		&		&		&		&		&		&				\\
\hline
\end{tabular}
\label{bica_confirmed}
\end{small}
\end{table*}

\begin{table*}\tabcolsep=9pt
\caption{Derived physical parameters of open cluster candidates.}
\begin{tabular}{l@{}c@{}c@{}c@{}c@{}c@{}l@{}l@{}c@{ }c@{ }r@{}} 
\hline
\multicolumn{1}{c}{Name}	&\multicolumn{1}{c}{$\alpha$(J2000)} &\multicolumn{1}{c}{$\delta$(J2000)}	&\multicolumn{1}{c}{Dist$_{\rm phot}$}	&	\multicolumn{1}{c}{Dist$_{\rm Gaia}$}	&	 \multicolumn{1}{c}{E($J$$-$$K_S$)} &	\multicolumn{1}{c}{Mass} &	\multicolumn{1}{c}{Age} &	\multicolumn{1}{c}{$\mu_\alpha\cos\delta$} &	\multicolumn{1}{c}{$\mu_\delta$} &	\multicolumn{1}{c}{$\rm [Fe/H]$} \\
\multicolumn{1}{c}{}	&	\multicolumn{1}{c}{$^\circ$} &	\multicolumn{1}{c}{$^\circ$}& \multicolumn{1}{c}{kpc} & \multicolumn{1}{c}{kpc} & \multicolumn{1}{c}{mag} & \multicolumn{1}{c}{M$_\odot$}	&	\multicolumn{1}{c}{Myr}  & \multicolumn{1}{c}{mas yr$^{-1}$} & \multicolumn{1}{c}{mas yr$^{-1}$}
&	\multicolumn{1}{c}{dex}	\\
\hline	
Bica\,372	&	112.7083 &	 $-$19.2572	&	3.1$\pm$0.27	&	3.1$\pm$1.6	&	0.57$\pm$0.04	&	2990$\pm$511	&	20$\pm$3	&	-1.48$\pm$0.88	&	1.68$\pm$1.20	&	-0.70$\pm$0.14	\\
Bica\,376	&	112.9625 &	 $-$19.2964	&	2.3$\pm$0.05	&	2.4$\pm$1.1	&	0.19$\pm$0.04	&	865$\pm$343	&	275$\pm$13	&	-2.00$\pm$0.53	&	2.34$\pm$0.51	&	0.07$\pm$0.02	\\
Bica\,377	&	112.9958	& $-$19.2986	&	3.8$\pm$0.31	&	2.3$\pm$1.2	&	0.43$\pm$0.06	&	5290$\pm$2160	&	375$\pm$13	&	-1.70$\pm$1.12	&	1.99$\pm$1.10	&	0.10$\pm$0.02	\\
Bica\,378	&	112.9833	& $-$19.3292	&	2.2$\pm$0.02	&	2.5$\pm$1.2	&	0.18$\pm$0.04	&	320$\pm$109	&	250$\pm$13	&	-1.83$\pm$0.62	&	2.10$\pm$0.84	&	-0.07$\pm$0.03	\\
Bica\,387	&	120.6375	& $-$34.5042	&	2.9$\pm$0.14	&	3.0$\pm$1.4	&	0.27$\pm$0.04	&	1120$\pm$442	&	350$\pm$25	&	-2.29$\pm$0.56	&	2.84$\pm$0.44	&	0.20$\pm$0.05	\\
Bica\,405	&	140.6500	& $-$51.9400	&	3.2$\pm$0.03	&	3.1$\pm$0.6	&	0.44$\pm$0.02	&	1640$\pm$215	&	650$\pm$25	&	-4.12$\pm$0.63	&	3.54$\pm$0.61	&	0.16$\pm$0.02	\\
Bica\,408	&	141.3833 &	 $-$52.1133	&	3.7$\pm$0.52	&	3.5$\pm$1.5	&	0.57$\pm$0.06	&	3620$\pm$1440	&	30$\pm$3	&	-3.70$\pm$0.85	&	3.12$\pm$0.74	&	0.10$\pm$0.02	\\
Bica\,422	&	159.6250	& $-$58.3133	&	2.0$\pm$0.03	&	1.6$\pm$0.9	&	0.13$\pm$0.04	&	884$\pm$308	&	500$\pm$25	&	-6.17$\pm$1.45	&	2.81$\pm$1.38	&	-0.23$\pm$0.10	\\
Cmg\,218	&	126.5792 &	 $-$40.8108	&	0.8$\pm$0.07	&	1.3$\pm$0.7	&	0.24$\pm$0.06	&	538$\pm$165	&	175$\pm$13	&	-2.45$\pm$0.78	&	3.18$\pm$1.56	&	-0.58$\pm$0.32	\\
Cmg\,398	&	149.5917 &	 $-$56.3703	&	2.0$\pm$0.21	&	2.3$\pm$0.9	&	0.53$\pm$0.04	&	530$\pm$259	&	70$\pm$10	&	-3.47$\pm$0.45	&	3.39$\pm$0.65	&	0.16$\pm$0.02	\\
Cmg\,405	&	149.7500 &	 $-$56.8617	&	2.7$\pm$0.36	&	2.3$\pm$1.1	&	0.48$\pm$0.04	&	1570$\pm$578	&	175$\pm$25	&	-5.23$\pm$0.82	&	3.51$\pm$1.35	&	0.18$\pm$0.02	\\
Cmg\,406	&	149.8125 &	 $-$56.9128	&	2.6$\pm$0.46	&	2.3$\pm$1.0	&	0.48$\pm$0.06	&	2100$\pm$844	&	60$\pm$5	&	-5.59$\pm$1.28	&	3.59$\pm$1.60&	0.27$\pm$0.02	\\
DBSB 9	&	112.9458	& $-$19.3672	  &	2.1$\pm$0.15	&	2.1$\pm$0.6	&	0.25$\pm$0.04	&	1340$\pm$400	&	600$\pm$50	&	-2.14$\pm$0.26	&	2.49$\pm$0.60	&	0.16$\pm$0.02	\\
Maj\,107	&	127.3083	& $-$41.1800	&	2.8$\pm$0.07	&	2.9$\pm$1.1	&	0.45$\pm$0.04	&	1280$\pm$397	&	45$\pm$3	&	-2.79$\pm$0.32	&	3.21$\pm$0.52	&	0.07$\pm$0.03	\\
Teu\,215	&	116.3750	& $-$20.8306	&	2.2$\pm$0.06	&	2.2$\pm$0.9	&	0.12$\pm$0.04	&	12000$\pm$3520\,\,\,	&	2500$\pm$120\,\,\,	&	-1.81$\pm$0.79	&	1.95$\pm$1.24	&	-0.23$\pm$0.05	\\
CL\,204	&	128.4500 &	 $-$44.4483	  &	0.7$\pm$0.01	&	3.0$\pm$1.2	&	0.18$\pm$0.04	&	397$\pm$59	&	7500$\pm$120\,\,\,	&	-2.71$\pm$0.42	&	3.40$\pm$0.71	&	-1.28$\pm$0.05	\\
CL\,207	&	145.6292 &	 $-$52.4306 	&	2.5$\pm$0.03	&	2.8$\pm$1.2	&	0.55$\pm$0.02	&	1030$\pm$119	&	500$\pm$25	&	-4.42$\pm$0.57	&	3.53$\pm$0.50	&	0.23$\pm$0.02	\\
CL\,208	&	148.5833 &	 $-$56.4233	  &	2.8$\pm$0.05	&	2.9$\pm$1.2	&	0.51$\pm$0.02	&	1960$\pm$231	&	200$\pm$13	&	-4.74$\pm$0.92	&	3.36$\pm$0.62	&	0.20$\pm$0.02	\\
\hline
\end{tabular}
\label{oc_known}
\end{table*}

\newpage

\begin{figure}
\includegraphics[width=15cm]{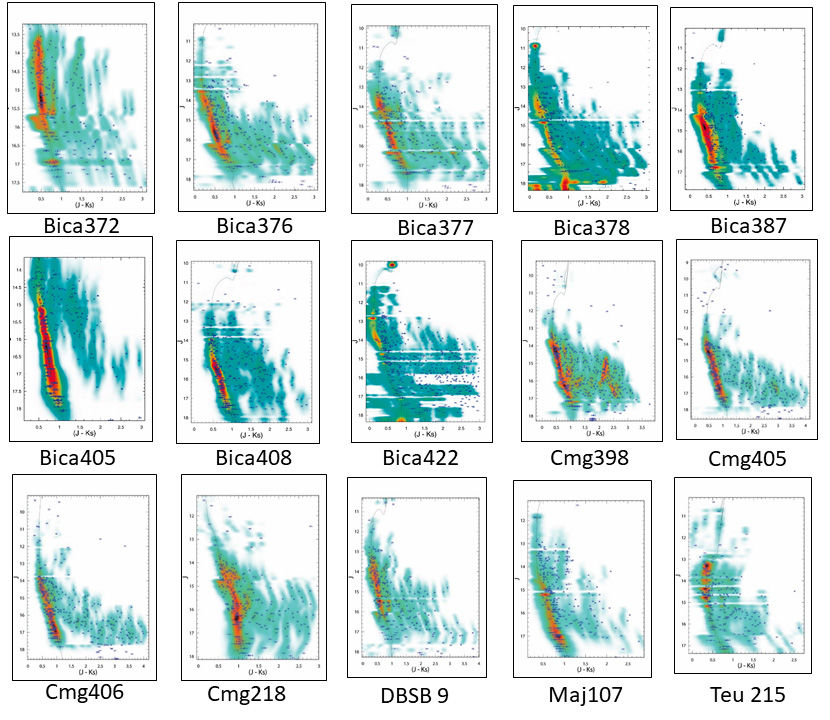}
\caption{VVVX ($(J-K_{s}), J$) Hess color magnitude diagrams  for some known open clusters from \citet{Bica19} catalog.} 
\label{cmd_nat}
\end{figure}

\bsp	

\label{lastpage}

\end{document}